\documentclass[twocolumn]{aastex63}
\usepackage{amssymb,amsmath}
\usepackage{graphicx}
\usepackage{xcolor,natbib}
\usepackage{multirow}
\usepackage[T1]{fontenc}
\usepackage{ae,aecompl}
\usepackage{newtxtext, newtxmath}
\usepackage{float}
\usepackage{subfigure}
\usepackage{longtable}
\usepackage{enumitem}
\usepackage{longtable,listings}
\usepackage[flushleft]{threeparttable}
\usepackage{parcolumns}
\def\oiii{[O~{\sc iii}]\ }
\def\nii{[N~{\sc ii}]\ }
\def\sii{[S~{\sc ii}]\ }
\def\oii{[O~{\sc ii}]$\lambda3727$\AA\ }

\def\o3c{[O~{\sc iii}]$_c$}
\def\o3b{[O~{\sc iii}$_B$]}
\accepted{\today}
\submitjournal{ApJ}

\shorttitle{CLQSO SDSS J2241}
\shortauthors{ZHANG}

\begin{document}

\title{The bluest changing-look QSO SDSS J224113-012108}

\correspondingauthor{XueGuang Zhang}
\email{xgzhang@njnu.edu.cn}
\author{XueGuang Zhang$^{*}$}
\affiliation{School of Physics and technology, Nanjing Normal University, No. 1, Wenyuan Road, Nanjing, 210023, P. R. China}




\begin{abstract} 
	In this manuscript, we report a new changing-look QSO (CLQSO) SDSS J2241 at $z=0.059$. 
Based on the multi-epoch SDSS spectra from 2011 to 2017, the flux ratio of broad H$\alpha$ 
to broad H$\beta$ has been changed from 7\ in 2011 to 2.7\ in 2017, leading SDSS J2241 with 
spectral index $\alpha_\lambda\sim-5.21\pm0.02$ ($\lambda< 4000$\AA) in 2017 to be so-far 
the bluest CLQSO. Based on the SDSS spectrum in 2011, the host galaxy contributions with 
stellar velocity dispersion $\sim86{\rm km/s}$ can be well determined, leading to the M-sigma 
relation expected central BH mass $\sim3\times10^6{\rm M_\odot}$. However, through properties 
of the broad H$\alpha$, the virial BH mass is $\sim10^8{\rm M_\odot}$, about two magnitudes 
larger than the mass through the M-sigma relation. The different BH masses through different 
methods indicate SDSS J2241 is one unique CLQSO. Meanwhile, the long-term photometric light 
curve shows interesting variability properties, not expected by DRW process commonly applied 
in AGN but probably connected to a central TDE. Furthermore, based on continuum emission 
properties in 2017 with no dust obscurations, only considering the moving dust clouds cannot 
be preferred to explain the CLQSO SDSS J2241, because the expected intrinsic reddening 
corrected continuum emissions were unreasonably higher than the unobscured continuum emissions 
in 2017.
\end{abstract}


\section{Introduction}

	Changing-look active galactic nuclei (CLAGN) have been studied for more than 
four decades, since the first CLAGN NGC 7603 reported in \citet{to76} with its broad 
H$\beta$ becoming much weaker in one year. Until now, there are about 40 CLAGN reported 
in the literature, according to the basic properties that spectral types of AGN are 
changing between type-1 AGN (apparent broad Balmer emission lines and/or Balmer 
decrements near to the theoretical values) and type-2 AGN (no apparent broad Balmer 
emission lines and/or Balmer decrements much different from the theoretical values). 
There are so-far dozens of CLAGN reported in the literature. 

     \citet{cr86} have reported the CLAGN Mrk1086 with its type changed from type-1.9 
to type-1\ in 4 years, and more recent results on the CLAGN Mrk 1018 can be found in 
\citet{mh16}. \citet{sb93} have reported the CLAGN NGC 1097 with the detected Seyfert 1 
nucleus which had previously shown only LINER characteristics. \citet{aj99} have reported 
the CLAGN NGC 7582 with the transition toward a type 1 Seyfert experienced by the classical 
type 2 Seyfert nucleus. \citet{eh01} have reported the CLAGN NGC 3065 with the new 
detected broad Balmer emission lines. \citet{dd14} have reported the CLAGN Mrk 590 with 
its type changed from Seyfert 1\ in 1970s to Seyfert 1.9\ in 2010s. \citet{sp14} have 
reported the CLAGN NGC 2617 classified as a Seyfert 1.8 galaxy in 2003 but as a Seyfert 1 
galaxy in 2013. \citet{la15} have reported the CLAGN SDSS J0159 classed as a type-1 AGN 
in 2000 but as a type-1.9 AGN in 2010. More recently, \citet{mr16} have reported ten 
CLAGN with variable and/or changing-look broad emission line features, and \citet{gh17} 
have reported the CLAGN SDSS J1554 with its type changed from type-2 to Type-1\ in 12 
years, and \citet{rf18} have reported two new changing-look quasar SDSS 
J1100-0053 through about 20years-long spectroscopic variabilities in different wavelength 
bands, and \citet{sm18} have reported a new changing-look quasar, WISE J1052+1519, found 
by identifying highly mid-IR variabilities, and \citet{yw18} have reported a sample of 
21 CLAGN with the appearance or the disappearance of broad Balmer emission lines within 
a few years.

	In order to explain the nature of CLAGN, different models have been proposed, 
such as the dynamical movement of dust clouds well discussed in \citet{em12}, the common 
variations in accretion rates well discussed in \citet{em14}, the variations in accretion 
rates due to transient events as discussed in \citet{er95, bn17}, and the 
magnetic torques in the free-fall region inside the ISCO (innermost stable circular orbit) 
as discussed in \citet{rf18, sm18}, etc. There is so-far no clear conclusion on the 
physical mechanism of type transitions in the CLAGN. For applied models on 
dust obscurations or on variations of central accretion process, well expected different 
effects on spectral energy distributions (SEDs) could be applied to determine which model 
plays the key role in type transition. More interestingly, to detect bluer changing-look 
AGN at bright state could provide clear clues to rule out the applications of dust 
obscurations, if intrinsic SEDs after corrections of dust obscurations were unreasonably 
beyond the observed SEDs at much bright state, such as discussed results in the reported 
CLAGN SDSS J2241.

\begin{figure*}
\centering\includegraphics[width = 18cm,height=4cm]{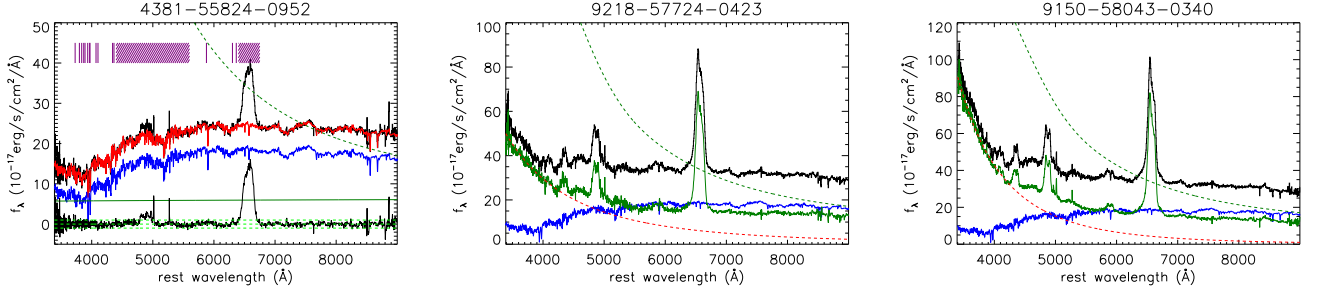}
\caption{Left panel shows the determined host galaxy contributions by the SSP method in 
the SDSS spectrum in 2011. From top to bottom, solid black line, solid red line, solid blue 
line, solid dark-green line, solid black line show the observed SDSS spectrum, the best 
descriptions to the observed SDSS spectrum, the determined host galaxy contribution, the 
determined power-law AGN continuum emissions and the pure line spectrum, respectively. The 
horizontal dashed green lines show $f_\lambda=0, \pm1$. The vertical lines in purple and 
the two areas filled by purple lines mark the emission features masked out, when the SSP 
method is applied. From left to right, the vertical purple lines point out \oii, H$\theta$, 
H$\eta$, [Ne~{\sc iii}]$\lambda3869$\AA, He~{\sc i}$\lambda3891$\AA, Ca K, 
[Ne~{\sc iii}]$\lambda3968$\AA, Ca H line, [S~{\sc ii}]$\lambda4070$\AA, H$\delta$, H$\gamma$, 
[O~{\sc iii}]$\lambda4364$\AA, He~{\sc i}$\lambda5877$\AA\ and [O~{\sc i}] doublet, 
respectively. The area filled by purple lines around 5000\AA\ shows the region masked out 
including the optical Fe~{\sc ii} lines, broad and narrow H$\beta$ and \oiii doublet, and 
the area filled by purple lines around 6550\AA\ shows the region masked out including the 
broad and narrow H$\alpha$, \nii and \sii doublets. Middle and right panels shows the SDSS 
spectra observed in 2016 and in 2017. In middle and right panels, solid black line, solid blue 
line, solid dark-green line and dashed red line show the SDSS spectrum, the host galaxy 
contribution shown in the left panel, the spectrum after subtraction of the host galaxy 
contribution and the power-law described continuum emissions with $\lambda<4000$\AA, respectively. 
The title of each panel shows the information of SDSS plate-mjd-fiberid. In each panel, 
the dashed dark-green line shows the reddening corrected intrinsic continuum emissions 
with $E(B-V)=0.8$, if the moving dust clouds was accepted.
}
\label{spec}
\end{figure*}

\begin{figure}
\centering\includegraphics[width = 8cm,height=11cm]{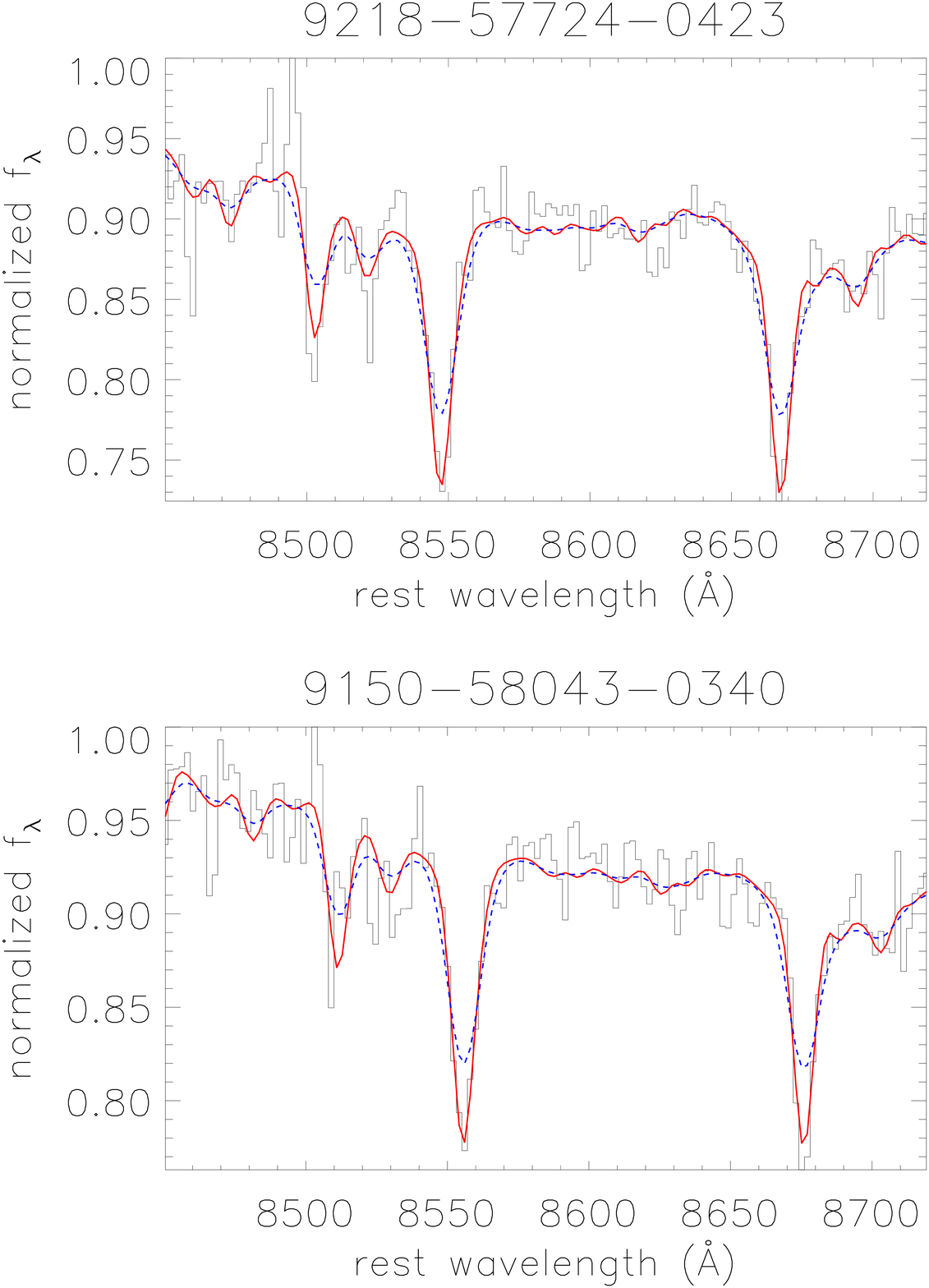}
\caption{Top and bottom panels show the best descriptions to the Ca~{\sc ii} triplets in the 
SDSS spectra in 2016 and in 2017. In each panel, solid black line and solid red line show the 
observed SDSS spectrum and the best descriptions, dashed blue line shows the 
compared component with broadening velocity about 100${\rm km/s}$.
}
\label{caii}
\end{figure}

\begin{figure*}
\centering\includegraphics[width = 18cm,height=12cm]{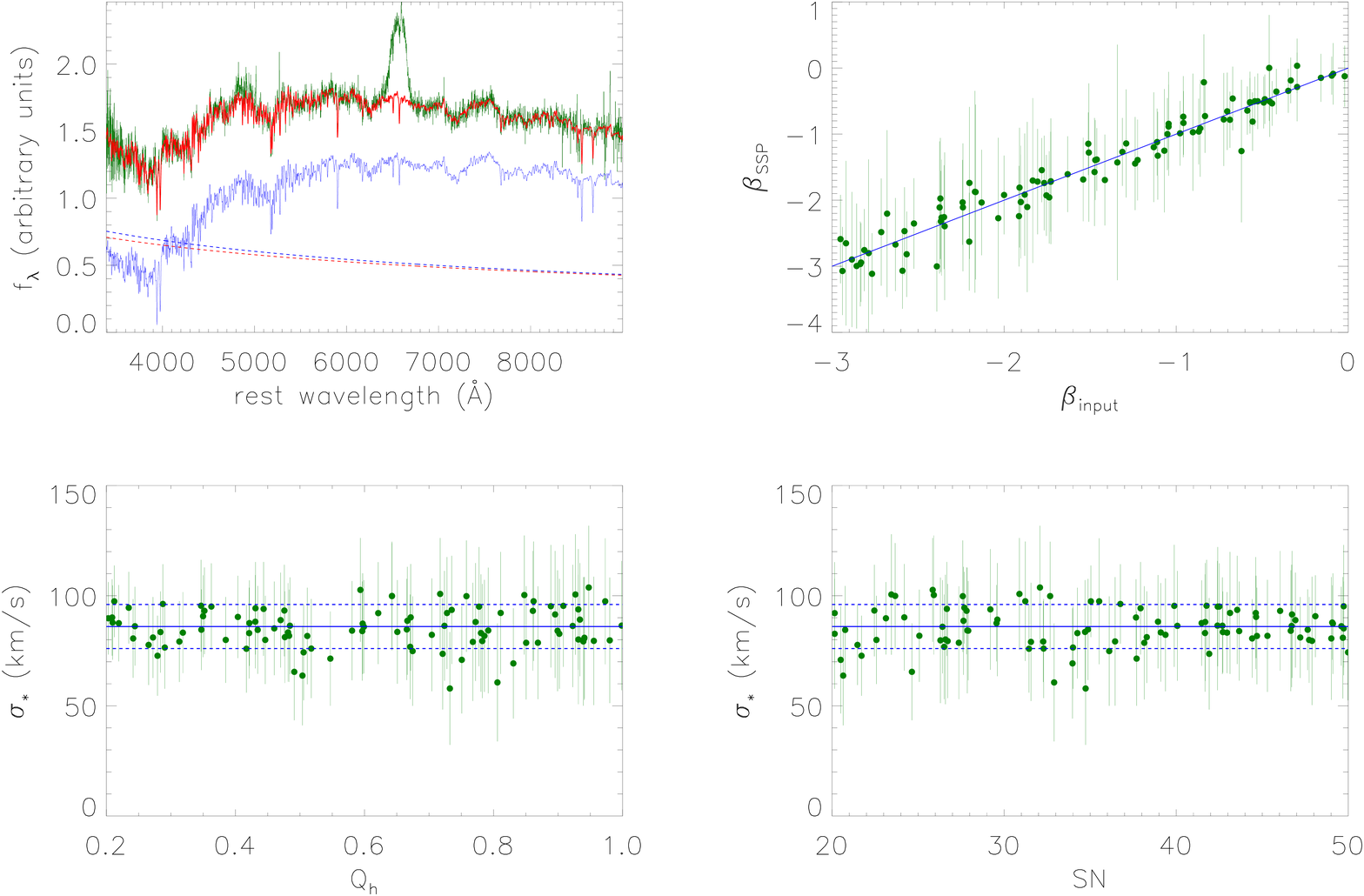}
\caption{Top left panel shows an example on the constructed synthetic spectrum with 
$Q_h=0.6$, $\beta=-0.57$ and $SN=49$ and the SSP method determined best descriptions to the synthetic 
spectrum. In the panel, solid blue line shows the stellar component $S_{lambda}$, dashed blue 
line shows the input power law component, solid dark green line shows the synthetic spectrum 
by $S_{lambda}$ plus the power law component and then adding a noise defined by SN=49, solid red 
line shows the SSP method determined best descriptions, dashed red line shows the SSP method 
determined power law component. Top right panel shows the correlation between the SSP method 
determined slope of the continuum emissions $\beta_{SSP}$ and the input value of $\beta_{input}$. 
In the panel, solid blue line shows $\beta_{SSP}=\beta_{input}$. Bottom two panels shown the 
dependence of the measured stellar velocity dispersions around 86${\rm km/s}$ on the $Q_h$ 
and $SN$. In each bottom panel, solid blue line shows $\sigma_*=86{\rm km/s}$, and dashed blue 
lines show $\sigma_*=86\pm10{\rm km/s}$.
}
\label{sp2}
\end{figure*}

 	Among the reported CLAGN, central BH masses have been estimated in SDSS J0159 and in 
SDSS J1554. The virial BH mass of SDSS J0159 have been well estimated as $1.6\times10^{8}{\rm M_\odot}$ 
in \citet{la15} and in \citet{zh19} under the Virialization assumptions to broad emission line 
regions \citet{pe04}, however, the measured stellar velocity dispersion is about $80{\rm km/s}$ 
SDSS J0159 in \citet{zh19}, leading to the much larger virial BH mass than the value through 
the M-sigma relation \citep{fm00, ge00, kh13, sg15}. Meanwhile, as the discussed results in SDSS 
J1554 in \citet{gh17}, the virial BH mass through properties of broad emission lines is about 
$2\times10^8{\rm M_\odot}$, and the measured stellar velocity dispersion is about $176{\rm km/s}$ 
leading to the M-sigma relation expected BH mass about $10^8{\rm M_\odot}$ well consistent 
with the virial BH mass. Therefore, as pointed out by \citet{yw18}, CLAGN can provide perfect 
cases to study the connection between AGN and their host galaxies, through properties in 
different states. And CLAGN having different BH masses from different methods could provide 
further interesting clues on the physical nature of type transition.

      In this manuscript, based on multi-epoch SDSS spectra, we report a new and so-far the 
bluest CLQSO SDSS J224113-012108 (=SDSS J2241) at $z=0.059$, of which both the host galaxy 
properties and the bright nuclei can be well determined during its type transitions,leading to 
quite different virial BH mass from the mass determined through the M-sigma relation (a detailed 
review can be found in \citet{kh13}), which will provide further clues on physical mechanism of 
type transitions in CLAGN. The manuscript is organized as follows. In section 2 and Section 3, 
the main results are shown on the spectroscopic properties and long-term photometric variability 
properties in SDSS J2241. In Section 4, the main discussions are given. Then, in Section 5, we 
show our main conclusions. And in the manuscript, we have adopted the cosmological parameters 
of $H_{0}=70{\rm km\cdot s}^{-1}{\rm Mpc}^{-1}$, $\Omega_{\Lambda}=0.7$ and $\Omega_{\rm m}=0.3$.

\begin{figure*}
\centering\includegraphics[width = 18cm,height=8cm]{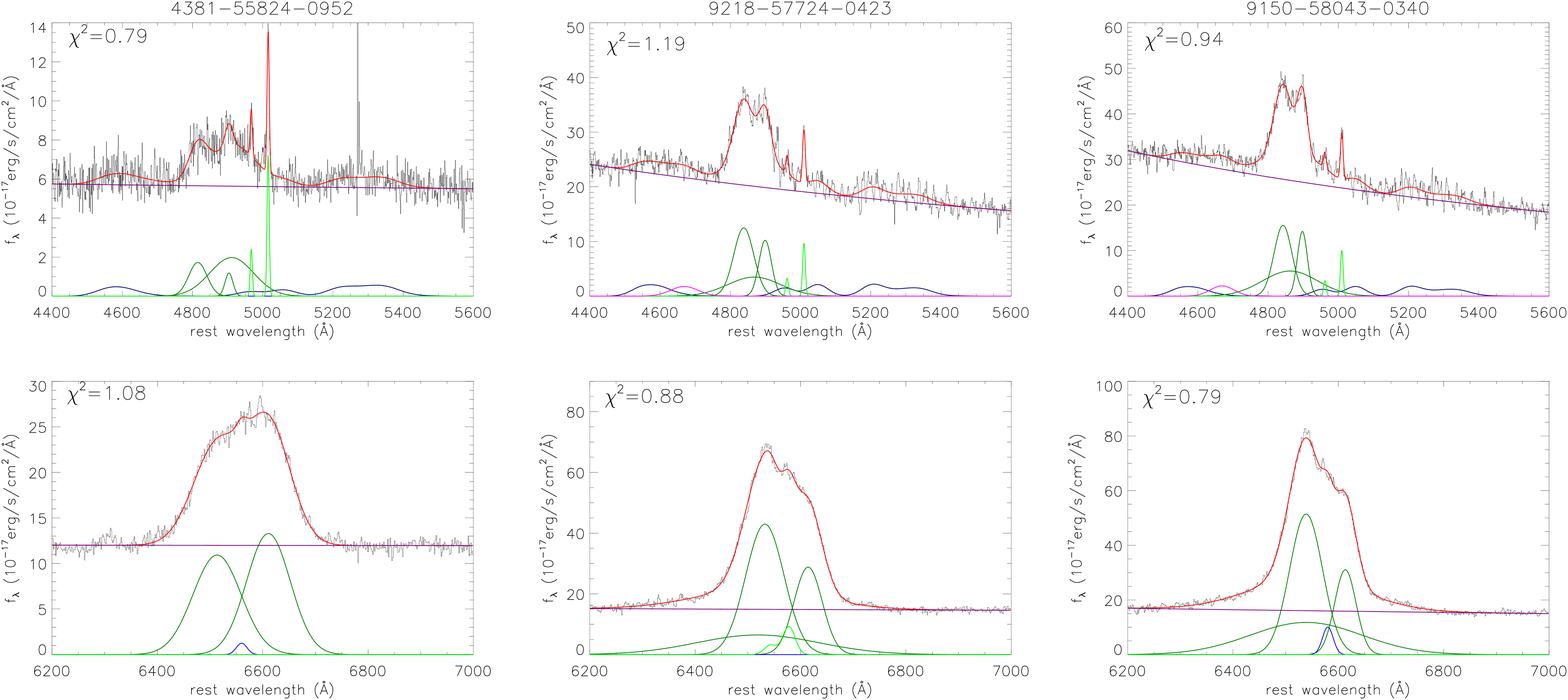}
\caption{The best descriptions to the emission lines around H$\alpha$ (bottom panels) and 
around H$\beta$ (top panels) in the SDSS spectrum in 2011 (left panels), in 2016 (middle panels) 
and in 2017 (right panels), after subtractions of host galaxy contributions. In each panel, solid 
black line shows the spectrum, solid red line shows the best fitted results, solid purple line 
shows the determined power law continuum emissions. In each top panel, solid dark-green lines 
show the determined three Gaussian components included in the broad H$\beta$, solid blue line 
shows the determined narrow H$\beta$, solid navy line shows the determined Fe~{\sc ii} lines, 
solid magenta line shows the determined He~{\sc ii} line, solid green lines show the determined 
[O~{\sc iii}] doublet. In each bottom panel, solid dark-green lines show the determined three 
Gaussian components included in the broad H$\alpha$, solid blue line shows the determined narrow 
H$\alpha$, solid green lines show the determine [N~{\sc ii}] doublet. The calculated $\chi^2$ 
values have been marked in top-left corner in each panel.
}
\label{line}
\end{figure*}

\begin{figure}
\centering\includegraphics[width = 8cm,height=5cm]{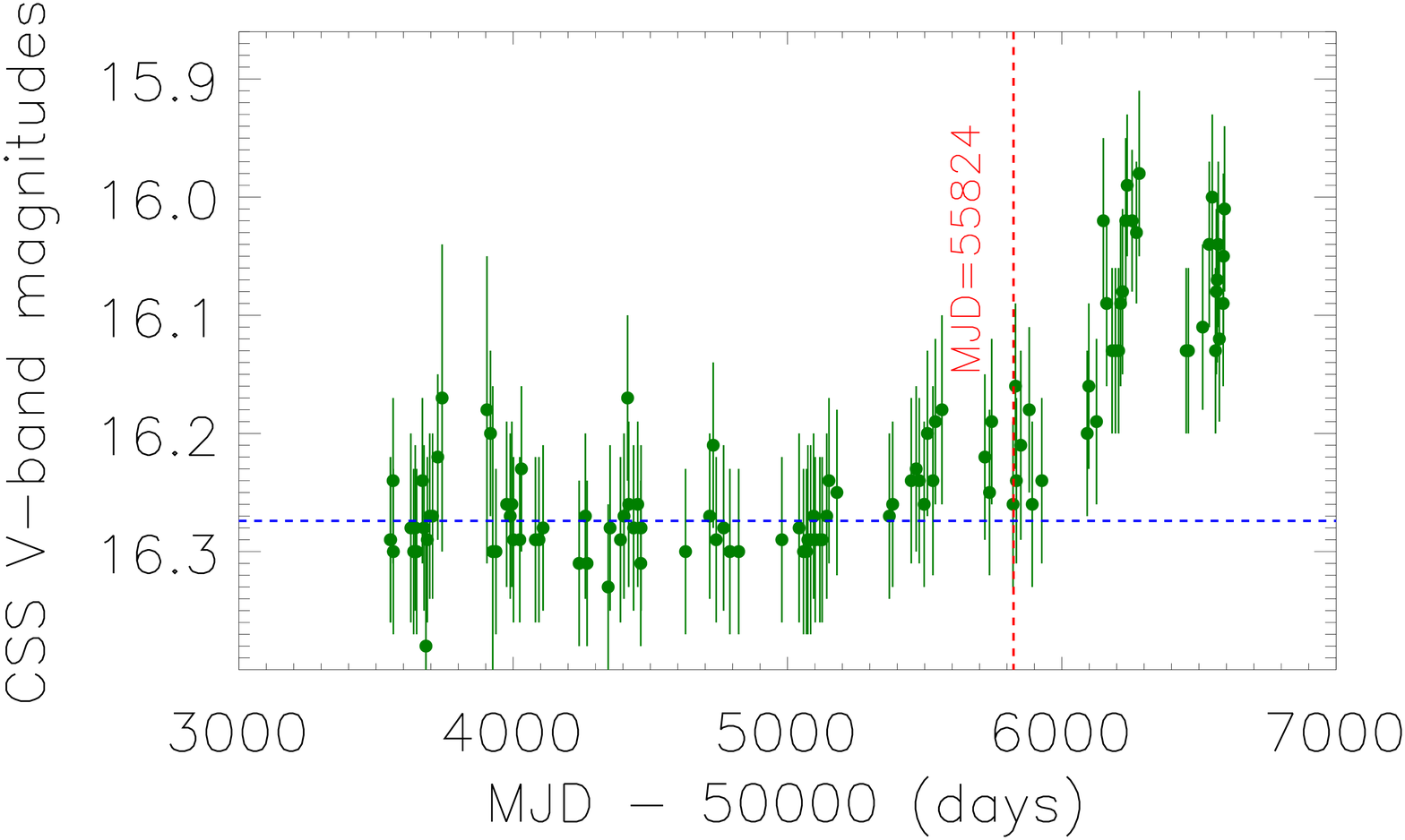}
\caption{Long-term CSS V-band photometric light curve of SDSS J2241. The vertical dashed 
red line marks the position MJD=55824 on which the spectrum SDSS 4381-55824-0952 observed 
and shown in the left panel of Fig.~\ref{spec}. The horizontal dashed blue line shows the 
mean apparent magnitude from the non-variable data points in the first 4.5 years.}
\label{lmc}
\end{figure}

\section{Spectroscopic Properties in SDSS J2241} 

    SDSS J2241, classified as a QSO in SDSS, has been observed in SDSS in September 2011, 
November 2016 and October 2017. Fig.~\ref{spec} shows the SDSS spectra with plate-mjd-fiberid 
as 4381-55824-0952, 9218-57724-0423 and 9150-58043-0340. It is clear that there are apparent 
contributions of host galaxy in the spectrum in 2011, and apparent blue AGN continuum 
emissions in the spectra in 2016 and 2017.

   The commonly accepted SSP (Simple Stellar Population) method has been well applied to 
determine the host galaxy contributions in the SDSS spectrum in 2011, similar as what we 
have done in \citet{zh14, zh16, zh19}. The more detailed descriptions on the SSP method can 
be found in \citet{bc03, ka03, cm05}. Here, the 39 simple stellar population templates 
from \citet{bc03} have been exploited, which can be used to well-describe the characteristics 
of almost all the SDSS galaxies as detailed discussions in \citet{bc03}. And, a power law 
component has been applied to describe the AGN continuum emissions, due to the apparent 
broad H$\alpha$ in the SDSS spectrum in 2011. After the emission lines being masked out, 
the observed SDSS spectrum can be well described by the broadened SSPs (the broaden velocity 
as the stellar velocity dispersion) plus the power law component through the Levenberg-Marquardt 
least-squares minimization technique, leading to $\chi2=SSR/Dof\sim1.3$ (where $SSR$ and $Dof$ 
as summed squared residuals and degree of freedom, respectively). Here, we have not only 
masked out all the narrow emissions lines with rest wavelengths between 3700\AA\ and 7000\AA~ 
with widths of about ${\rm 450km/s}$, mainly including [O~{\sc ii}]$\lambda3727$\AA, narrow 
Balmer lines, [O~{\sc iii}]$\lambda4364$\AA, [O~{\sc iii}] doublet, [O~{\sc i}] doublet, 
[N~{\sc ii}] doublet and [S~{\sc ii}] doublet, etc., but also masked out the optical 
Fe~{\sc ii} lines and the broad H$\alpha$ and broad H$\beta$, which have been shown in 
the left panel of Fig.~\ref{spec}. The best descriptions to the host galaxy contributions 
have been shown in the left panel of Fig.~\ref{spec}. The determined stellar velocity 
dispersion is about $\sigma_\star\sim87\pm5{\rm km/s}$, after considering $60{\rm km/s}$ 
as the mean SDSS instrument broadening velocity. And the determined power law AGN component 
has $f_\lambda\propto\lambda^{0.06\pm0.12}$ in the SDSS spectrum in 2011.

   Meanwhile, after subtractions of host galaxy contributions from the SDSS spectra in 
2016 and in 2017, the blue parts of the spectra with rest wavelength less than 4000\AA\ 
can be well described by $f_\lambda\propto\lambda^{-3.83\pm0.04}$ and 
$f_\lambda\propto\lambda^{-5.21\pm0.02}$ in 2016 and in 2017, respectively. Meanwhile, 
among the reported CLAGN in the literature, SDSS J1100-0053 well discussed in \citet{rf18} 
is also a bright quasar at its bright state, with the blue part of the spectra with rest 
wavelength less than 4000\AA\ described by $f_\lambda\propto\lambda^{-2.45}$. Therefore, 
among the reported CLAGN, SDSS J2241 is the bluest changing-look QSO at its bright state.

   Besides the SSP method determined $\sigma_\star$, the apparent Ca~{\sc ii} triplet 
around 8500\AA\ well detected in the SDSS spectra in 2016 and in 2017 can provide another 
method to estimate $\sigma_\star$. Based on the 1273 template stellar spectra with higher 
resolution about ${\rm 30~km/s}$ collected from the Indo-U.S. Coude Feed Spectral Library 
\citep{vg04}, the stellar velocity dispersion is re-measured based on the best descriptions 
to the Ca~{\sc ii} triplet through the Levenberg-Marquardt least-squares minimization 
technique, similar as what have been done in \citet{rw92, gh06}. Here, besides the broadened 
template stellar spectra, a three-order polynomial function has been applied to modify the 
continuum shape. The best descriptions have been shown Fig.~\ref{caii}, with the calculated 
$\chi2$ values around 0.8 and 0.6 for the results to the spectra in 2016 and in 2017. After 
considering the spectral resolution of the template stellar spectra and the SDSS instrument 
broadening velocity of $50{\rm km/s}$ around the Ca~{\sc ii} triplet, the re-measured 
stellar velocity dispersions are about $83\pm10{\rm km/s}$ and $86\pm12{\rm km/s}$ for 
the results to the spectra in 2016 and in 2017. The re-measured $\sigma_\star$ through 
the Ca~{\sc ii} triplets are well consistent with the determined value through the whole 
spectrum observed in 2011. Therefore, the mean value of $\sigma_\star\sim86\pm12{\rm km/s}$ 
is accepted as the stellar velocity dispersion of SDSS J2241.

     Before proceeding further, as what we have known that both the ratio $Q_h$ of the 
quasar luminosity to the host galaxy luminosity and the overall signal-to-noise ($SN$) 
of the spectrum have apparent effects on the measured stellar velocity dispersion of 
host galaxy of a quasar, the following two methods are applied to well demonstrate that 
the measured stellar velocity dispersion about 86$km/s$ is reliable enough in SDSS J2241. 
On the one hand, properties of Ca~{\sc ii} triplet are mainly considered as follows. 
Besides the best descriptions to the Ca~{\sc ii} triplet shown in Figure~\ref{caii}, 
a new component described by the same stellar template with the same strengthen factor 
but with broadened velocity of 100${\rm km/s}$ is also shown in the Figure~\ref{caii}. 
The quite difference between the best descriptions and the new component with broadened 
velocity of 100${\rm km/s}$ strongly indicate that the measured stellar velocity 
dispersion is well smaller than 100${\rm km/s}$ in SDSS J2241. On the other hand, 
properties of the whole spectra of SDSS J2241 are mainly considered as follows. Based 
on the best descriptions $S_{\lambda}$ to the stellar lights shown in left panel of 
Figure~\ref{spec}, a series of 100 synthetic spectra without effects of noise can be 
well constructed by the $S_{lambda}$ normalized by $S_{5100}=1$ plus a power law 
component $Q_h\times(\lambda/5100)^\beta$ ($Q_h$ as a random value from 0.2 to 1, 
$Q_h=0.4$ is the value for the results in 2011 shown in the left panel of Figure~\ref{spec}) 
which is applied to describe intrinsic AGN continuum emissions. Besides considering 
effects of $Q_h$, different values of $SN$ from 20 to 50 ($SN\sim40$ for the spectrum 
observed in 2011\ in SDSS J2241) are considering to add noise to the synthetic spectra. 
Then the similar SSP procedure is applied to measure the stellar velocity dispersions 
of the 100 synthetic spectra with considerations of both $Q_h$ and $SN$. Top left panel 
of Figure~\ref{sp2} shows one constructed synthetic spectrum (with $Q_h=0.6$, $\beta=-0.57$ 
and $SN=49$) and the determined best descriptions to the synthetic spectrum. Top right 
panel of Figure~\ref{sp2} shows the correlation between the SSP method determined 
slope of the continuum emissions $\beta_{SSP}$ and the input value of $\beta_{input}$, 
with mean value of $\beta_{SSP}/\beta_{input}\sim1.03$. Bottom panels of Figure~\ref{sp2} 
shown the dependence of measured stellar velocity dispersions around 86${\rm km/s}$ 
on the $Q_h$ and $SN$. It is clear that there are few effects of $Q_h$ and $SN$ on 
the measured stellar velocity dispersion. Therefore, the measured stellar velocity 
dispersion in SDSS J2241 is reliable enough. Then, the SSP method determined stellar 
lights can be well subtracted, in order to well measure properties of emission lines 
of SDSS J2241.

     After subtractions of host galaxy contributions, emission lines, especially around 
the broad H$\alpha$ (rest wavelength between 6200\AA\ and 7000\AA) and around the broad 
H$\beta$ (rest wavelength between 4400\AA\ and 5600\AA), can be well measured by the following 
model functions. Around H$\alpha$, due to double-peaked features of broad H$\alpha$ and quite 
weak narrow lines of [ O~{\sc i}] and [S~{\sc ii}] doublets, three broad Gaussian functions 
are applied to describe the broad H$\alpha$, and only three narrow Gaussian components are 
applied to describe the narrow H$\beta$ and [N~{\sc ii}] doublet, and a power law component 
is applied to describe the continuum emissions underneath the broad H$\alpha$. Around 
H$\beta$, three Gaussian components are applied to describe the double-peaked broad H$\beta$, 
one narrow Gaussian component is applied to describe the narrow H$\beta$, two Gaussian 
components are applied to describe the [O~{\sc iii}] doublet, one broad Gaussian component 
is applied to describe broad He~{\sc ii}, the Fe~{\sc ii} template discussed in \citet{kp10} 
is applied to describe the optical Fe~{\sc ii} lines and a power law component is applied 
to describe the continuum emissions underneath the broad H$\beta$. Then, through the 
Levenberg-Marquardt least-squares minimization technique, the emission lines can be well 
described by the model functions, which have been shown in Fig.~\ref{line}. 
The determined parameters have been listed in Table~1 for the broad Gaussian components 
in the broad Balmer lines. Before proceeding further, there is one point we should note. 
There are determined narrow emission lines shown in Fig.~\ref{line}, however, the determined 
fluxes of the narrow emission lines are smaller than 5 times of their corresponding flux 
uncertainties. Therefore, line parameters of the narrow emission lines are not listed in 
Table~1.

	Based on the measured line parameters of the broad Balmer lines, the flux ratios 
of broad H$\alpha$ to broad H$\beta$ are about 6.98, 2.86 and 2.79\ in 2011, 2016 and 2017, 
respectively. The apparent changes of Balmer decrements in six years lead SDSS J2241 to be 
a new CLAGN. Meanwhile, based on the power law component in the blue part of the spectrum shown 
in the bottom panel of Fig.~\ref{spec} with $f_\lambda\propto\lambda^{-5.21\pm0.02}$, 
SDSS J2241 is the bluest CLQSO among the reported CLAGN.

\begin{table*}
\caption{Line parameters of emission lines}
\begin{tabular}{lccccccccc}
\hline\hline
 & \multicolumn{3}{c}{parameters in 2011} & \multicolumn{3}{c}{parameters in 2016} & \multicolumn{3}{c}{parameters in 2017}  \\
\hline
Line    &   $\lambda_0$  &   $\sigma$  & flux &   $\lambda_0$  &   $\sigma$  & flux &   $\lambda_0$  &   $\sigma$  & flux \\
\hline
H$\alpha_{B1}$ & 6513.7$\pm$2.8 & 45.8$\pm$1.7 & 1259$\pm$73 &
		6522.1$\pm$3.4 & 112.8$\pm$4.3 & 2333$\pm$83 &
		6544.7$\pm$1.6 & 107.7$\pm$2.5 & 3528$\pm$75 \\

H$\alpha_{B2}$ &     &     &      &
	      6535.1$\pm$1.3   & 36.1$\pm$0.9 & 3851$\pm$180 &
	      6545.3$\pm$1.1 & 30.6$\pm$0.7  & 3740$\pm$161 \\

H$\alpha_{B3}$ &  6611.1$\pm$2.1 & 41.4$\pm$1.5 & 1373$\pm$71 & 
	6618.4$\pm$1.4   & 25.7$\pm$1.0 & 1799$\pm$90 & 
	6622.3$\pm$0.9 & 19.9$\pm$0.7  & 1572$\pm$68 \\
\hline

H$\beta_{B1}$ & 4816.6$\pm$4.9 & 28.1$\pm$4.1 & 91$\pm$52 &
	       4837.7$\pm$1.1  & 27.1$\pm$1.1 & 829$\pm$49 &
	       4848.2$\pm$0.9  & 26.3$\pm$1.2 & 1043$\pm$93 \\

H$\beta_{B2}$ & 4902.5$\pm$3.1  & 12.7$\pm$4.4 & 30$\pm$13 &
                4878.9$\pm$4.1  & 96.1$\pm$4.4 & 1537$\pm$87 & 
 		4871.9$\pm$5.8  & 94.9$\pm$7.7 &  1643$\pm$154 \\
H$\beta_{B3}$ & 4911.3$\pm$11.3 & 65.7$\pm$11.5 & 260$\pm$57 &
	        4900.8$\pm$0.9 & 17.3$\pm$0.9 & 425$\pm$32 & 
		4904.4$\pm$0.7 & 13.8$\pm$0.8 & 482$\pm$44 \\
\hline
\end{tabular}\\
{\bf Notice:} The second, third and fourth columns show the rest central wavelength in unit 
of \AA, the line width (second moment) in unit of \AA\ and the line flux in unit of 
$10^{-17}{\rm erg/s/cm^2}$ from the spectrum in 2011. The fifth to seventh columns show the 
parameters from the spectrum in 2016, and the last columns show the parameters from the 
spectrum in 2017. The suffix "B1", "B2" and "B3" represent the three Gaussian components 
sorted by rest central wavelength in the broad Balmer lines. 
\end{table*}

\section{Photometric Properties in SDSS J2241}

    Fig.~\ref{lmc} shows the CSS V-band photometric light curve of SDSS J2241 collected 
from the Catalina Sky Survey (CSS) \citep{dr09}. Apparently, the long-term variabilities 
are very interesting in more than 8 years from July 2005 to October 2013 (MJD from 53552 to 
56593). In the first 4.5 years with MJD from 53552 to 55261, there are no apparent variabilities 
in SDSS J2241. However since MJD=55261, the nucleus becomes brighter and brighter. Certainly, 
the long-term variability properties cannot be expected by the DRW (Damped Random Walk) 
process which has been proved to be a preferred modeling process to describe AGN intrinsic 
variability \citep{kbs09, koz10, ak13, zu13}. 

\section{Discussions}

\subsection{Virial BH mass and BH mass through M-sigma relation}

	Based on the measured line parameters, the second moments $\sigma_{\rm B}$ of broad 
H$\alpha$ are about $2980\pm270{\rm km/s}$, $3480\pm230{\rm km/s}$ and $3520\pm240{\rm km/s}$ 
in 2011, 2016 and 2017, respectively. And the FWHM (full width at half maximum) of broad 
H$\alpha$ are about $8494\pm770{\rm km/s}$, $6947\pm470{\rm km/s}$ and $6170\pm420{\rm km/s}$ 
in 2011, 2016 and 2017, respectively. The line luminosities of broad H$\alpha$ ($L_{\rm B}$) 
are about $(2.12\pm0.12)\times10^{41}{\rm erg/s}$, $(6.42\pm0.28)\times10^{41}{\rm erg/s}$ 
and $(7.11\pm0.25)\times10^{41}{\rm erg/s}$ in 2011, 2016 and 2017, respectively. It is clear 
that FWHMs being decreased with line luminosity being increased is consistent with the 
expected results under the Virialization assumption to broad Balmer lines, however varying 
of $\sigma_{\rm B}$ does not. Therefore, rather than the $\sigma_{\rm B}$, the FWHM is 
preferred to estimate the virial BH masses in SDSS J2241, through the following equation 
well discussed in \citet{gh05} after considering the more recent improved R-L relation 
in \citet{ben13}
\begin{equation}
\frac{M_{\rm BH}}{\rm M_\odot}=2.4\times10^6(\frac{L_{\rm B}}{\rm 10^{42}erg/s})^{0.47}(\frac{FWHM}{\rm 1000km/s})^{2.06}
\end{equation}
Then, the virial BH masses are about $(9.5\pm2.5)\times10^{7}{\rm M_\odot}$, 
$(10.5\pm2.9)\times10^{7}{\rm M_\odot}$ and $(8.7\pm1.9)\times10^{7}{\rm M_\odot}$ in 2011, 
2016 and 2017, respectively. Then, the mean value $(9.6\pm2.4)\times10^{7}{\rm M_\odot}$ is 
accepted as the virial BH mass of SDSS J2241. 

	Meanwhile, based on the measured stellar velocity dispersion $\sigma_\star=86\pm12{\rm km/s}$, 
the M-sigma relation expected BH mass is about $2.8\times10^6{\rm M_\odot}$ through the M-sigma 
relation discussed in \citet{kh13} and in \citet{zh19}. Therefore, the virial BH mass is about 
two magnitudes larger than the BH mass through the M-sigma relation. The larger virial BH mass 
could provide further clues on special dynamical properties of BLRs. Furthermore, as the shown 
results in Fig.~\ref{line}, there are much different line profiles of broad H$\alpha$ and broad 
H$\beta$: apparent double peaks in broad H$\beta$ but one peak plus one shoulder in broad H$\alpha$ 
(especially in 2016 and 2017). We can not find natural explanations on the different line profiles, 
but the different line profiles can indicate there could be non-keplerian components in broad 
Balmer emission regions. Further efforts should be necessary to check the variability properties 
of broad emission lines.

\subsection{Moving dust clouds?}

     In the CLQSO SDSS J2241, if moving dust clouds was applied to explain the changes of 
Balmer decrement from 6.98\ in 2011 to 2.79\ in 2017, the $E(B-V)=0.8$ could be estimated, 
assuming the theoretical Balmer decrement to be 2.80. Then, intrinsic reddening-corrected 
continuum emissions could be checked in 2011, and shown as dashed dark-green lines in 
Fig.~\ref{spec}. It is clear that considering the moving dust clouds to explain the changes 
of Balmer decrements, the expected intrinsic continuum emissions could be unreasonably 
larger than the continuum emissions in 2017 with no dust obscurations. Therefore, the 
moving dust clouds can not be applied in SDSS J2241. Although, we can not totally ruled out 
the moving dust clouds to explain the CLQSO SDSS J2241, the spectra of redward of broad 
H$\alpha$ have tiny variations of spectral slope against the expected results by varying 
obscurations.

    Actually, besides the well shown properties of variations of continuum emissions in 
SDSS J2241, there are convincing methods applied to rule out the moving dust clouds in CLAGN, 
such as the timescale arguments as well discussed in \citet{la15, rc16, rf18} and the 
expected profiles of the broad emission lines in \citet{ra16}. Then, the two methods are 
applied in SDSS J2241 as follows. 

    On the one hand, we consider the timescale arguments as follows. As discussed in 
\citet{la15, rc16}, etc, it is necessary and interesting to check whether the timescale 
$t_{cross}\sim0.07r_{orb}^{1.5}M_{8}^{-0.5}\arcsin(r_{src}/r_{orb})$ is short 
enough for foreground moving dust clouds in a bound orbit around the central BH in 
front of the continuum emission source and BLRs, where $r_{orb}$ is radius of the foreground 
moving dust clouds, $M_8$ is the BH mass in units of $10^{8}{\rm M_\odot}$, and $r_{src}$ 
is the true size of the BLRs. Commonly, there are no accurate values of the parameters of 
$r_{orb}$, $M_8$ and $r_{src}$. However, it is not difficult to estimate a minimum timescale 
of $t_{cross}$, after considerations of the estimated size of central BLRs. Here, at the 
bright state of SDSS J2241 with continuum luminosity at 5100\AA~ about 
$1.03\times10^{43}{\rm erg/s}$ in 2017, the size of central BLRs is estimated as 
$R_{BLRs}\sim11{\rm light-days}$ through the improved empirical R-L relation reported 
in \citet{ben13}. Similar as what have been discussed in \citet{la15, rc16}, etc, 
assuming $r_{orb}\ge k_o\times R_{BLRs}$ and $r_{src}\sim k_s\times R_{BLRs}$, the 
cross timescale can be estimated as 
\begin{equation}
t_{cross}\sim2.6M_8^{-0.5}k_o^1.5\arcsin(k_s/k_o){\rm years}
\end{equation}.
The estimated cross timescale sensitively depends on both the central BH mass and the 
parameters of $k_s$ and $k_o$. If we accepted that $k_o=k_s=3$ as what have been set in 
\citet{la15, rc16} and accepted BH mass about $10^8{\rm M_\odot}$ (the virial BH mass), 
the minimum cross timescale is about 21years. Moreover, if central BH mass 
$2.8\times10^6{\rm M_\odot}$ through the M-sigma relation is accepted, the minimum cross 
timescale should be 125years. It is clear that the commonly estimated cross timescale 
is quite longer than the observed transition time in SDSS J2241. Therefore, the timescale 
arguments can also be well applied to rule out the moving dust clouds in SDSS J2241. 

   On the other hand, we discuss the arguments on broad line profiles as follows. As well 
discussed in \citet{ra16}, the applied moving dust clouds have apparent effects on line 
profiles of broad emission lines, indicating that broad line emissions from the outer 
lower-velocity emission regions of the broad-line region are attenuated more than emissions 
from the inner emission regions, leading to broad optical emission lines (such as the broad 
Balmer lines) are broader in the dim state. However, not similar as the findings in SDSS 
J2336 shown in Figure~4\ in \citet{ra16}, the line profiles of broad H$\alpha$ are more 
complicated in SDSS J2241. And as the results shown above, if considering second moment 
as the line width, the broad H$\alpha$ become broader from 2011 to 2017, however, if 
considering FWHM as the line width, the broad H$\alpha$ become narrower 2011 to 2017. 
Therefore, properties of line profiles of broad emission lines can not provide clear 
clues to rule out the scenario of moving dust clouds in SDSS J2241.

    Based on the discussions, rather than moving dust clouds, accretion rate variations are 
preferred to explain the changing-look QSO SDSS J2241.

\subsection{Accretion rate variations?}

	TDEs (tidal disruption events) can be considered as one of optimal choices 
leading to variations of accretion rate in central regions of galaxies. More detailed 
discussions on TDEs can be found in \citet{re88, gr13, gm14, st18}. And there are more 
than 80 TDEs detected and reported (see detailed information in \url{https://tde.space/}), 
strongly supporting TDEs as better indicators to massive BHs and corresponding BH accreting 
systems. The following points could provide clues on a central TDE candidate in SDSS J2241. 
First, the smooth photometric variability trend is preferred to be connected with a TDE 
rather than common AGN variabilities. Second, there are broad He~{\sc ii} lines in 2016 
and 2017 but quite weak He~{\sc ii} line in 2011, which is characteristic of TDE spectra. 
Third, the broad Balmer line profiles have apparent variations from 2011 to 2017, possibly 
due to structure evolutions of TDE debris. Fourth, virial BH mass is not consistent with 
the mass through the M-sigma relation, possibly similar as the case in the TDE candidate 
of SDSS J0159 \citep{zh19}. Certainly, further efforts are necessary to identify whether 
there is a central TDE in SDSS J2241.

   Moreover, \citet{rc16} have reported a CLAGN SDSS J1101 of which the smooth decay of 
its long-term photometric variabilities can be well consistent with a slope of $-5/3$ 
(see the results shown in Figure~6\ in \citet{rc16}) which is the value expected by the 
theoretical TDE model \citep{gr13}. However, after considering the longer bright state 
lasting at least 6years and more massive broad emission line regions which can not 
constructed by TDEs debris from a solar-like star and the peculiar properties of narrow 
emission lines in SDSS J1101 with stronger [O~{\sc iii}] luminosity, \citet{rc16} have 
concluded that the TDE scenario is not preferred to explain the type transition of SDSS 
J1101. Similar discussions to rule out a TDE scenario can also be found in \citet{ra16}. 
The considerations above on TDEs scenario in \citet{rc16, ra16} should also be well 
considered in SDSS J2241 by the following three points. First and foremost, it is 
unfortunate that the collected photometric light curve is not long enough to detect 
whether the bright state lasts longer as the one shown in SDSS J1101. Besides, if broad 
emission lines in SDSS J2241 were also considered from emission regions totally constructed 
by TDEs debris as the ones discussed in SDSS J1101 in \citet{rc16}, the conclusion 
should be also clear to rule out the TDE scenario applied in the type transition in 
SDSS J2241. However, once accepted that only part of broad line emissions of SDSS J2241 
have contributions from expected TDEs debris (a large part of broad line emissions from 
central normal BLRs connected with intrinsic AGN, i.e., a probable TDE in AGN), the 
strong broad line features should be not robust evidence to rule out the TDE scenario in 
type transition of SDSS J2241. Last but not the least, the [O~{\sc iii}] luminosity of 
SDSS J2241 can be measured as $7.7\times10^{39}{\rm erg/s}$, at least two magnitudes 
smaller than the listed values in \citet{ra16}. And moreover, not similar as the strong 
narrow emission lines measured at the dime states in the CLAGN in \citet{ra16, rc16}, 
there are no narrow Balmer emission lines clearly detected at the dime state in SDSS 
J2241 as the results shown in Figure~\ref{line}. The weak narrow Balmer emission lines 
in SDSS J2241 are similar as the optical spectroscopic emission line properties of 
the reported TDEs, which could be treated as clues to support a central TDE in SDSS 
J2241. In one word, as discussed above, further efforts are necessary to identify 
whether there is a central TDE in SDSS J2241.

\section{Conclusions}

   Finally, we give our main conclusions as follows. Based on the SDSS spectra observed 
in 2011, 2016 and 2017, the flux ratio of broad H$\alpha$ to broad H$\beta$ is changed 
from 7\ in 2011 to 2.7\ in 2017\ in the QSO SDSS J2241, leading SDSS J2241 to be a new 
CLQSO. And moreover, after considering host galaxy contribution, the spectral index 
$\alpha_\lambda\sim-5.21\pm0.02$ with $\lambda<4000$\AA\ can be well determined in 2017, 
leading SDSS J2241 to be so-far the bluest CLQSO. Meanwhile, based on the determined 
stellar velocity dispersion about $86{\rm km/s}$ and the properties of broad H$\alpha$, 
the virial BH mass in SDSS J2241 is about two magnitudes larger than the BH mass through 
the M-sigma relation, not similar as the clear case in the CLAGN SDSS J1554. Meanwhile, 
the long-term photometric variability curve shows interesting smooth slowly gradual 
changing properties, not expected by DRW process commonly applied in common AGN. Moreover, 
based on the properties of continuum emissions in 2017 with no dust obscurations, only 
considering the moving dust clouds cannot be preferred to explain the CLQSO SDSS J2241, 
because the expected intrinsic reddening corrected continuum emissions in 2011 were 
unreasonably higher than the unobscured continuum emissions in 2017. The accretion rate 
variations can be preferred, and the probable accretion rate variability could be 
probably due to a TDE in SDSS J2241.

\section*{Acknowledgements}
Zhang gratefully acknowledge our referee for reading our manuscript carefully 
and patiently, and gratefully acknowledge the our referee for giving us constructive 
comments and suggestions to greatly improve our paper. Zhang gratefully acknowledges 
the kind support of Starting Research Fund of Nanjing Normal University and from the 
financial support of NSFC-11973029. This manuscript has made use of the data from the SDSS 
projects. The SDSS-III web site is http://www.sdss3.org/. SDSS-III is managed by the 
Astrophysical Research Consortium for the Participating Institutions of the SDSS-III Collaboration.


\end{document}